\documentclass[traditabstract]{aa} 
\usepackage[pdftex]{graphicx}
\usepackage{txfonts}
\begin{document}
\title{Multiple outflows from the luminous YSO IRAS~07422-2001}


\author{Watson P. Varricatt}

\institute{Joint Astronomy Centre,
   	   660 N. Aohoku pl., Hilo, HI-96720, USA\\
          \email{w.varricatt@jach.hawaii.edu}
             }

\date{Received 03 October 2011; Accepted 14 February 2012}

 
\abstract
{The luminous Young Stellar Object (YSO) IRAS~07422-2001 
is studied in the
infrared.  We discover star forming activity in embedded 
clusters located in a cloud detected at mid-IR wavelengths 
in emission. Multiple outflows are discovered from these
clusters in the H$_2$ ro-vibrational line at 2.122\,$\mu$m.  
We detect at least six outflows from the cluster associated 
with the IRAS source and another outflow from a source 
located in a cluster detected $\sim$2.7\,arcmin NE of the 
IRAS source.  Additional star formation is taking place
in two other cluster candidates within the cloud.  Three 
of the YSOs in the cluster associated with the IRAS 
source are detected at 11.2\,$\mu$m at an angular 
resolution of $\sim$0.8\arcsec.  We have a tentative
detection of a circumstellar disk in this cluster, 
seen as an extinction lane in the $J$ and $H$-band
images.  The spectral energy distributions (SEDs) of 
the dominant YSOs in the cluster associated with the 
IRAS source and in the NE cluster are studied using 
radiative transfer models and the properties 
of the YSOs are estimated.  The YSO associated with the
IRAS source is  probably in a very early Class I stage of 
formation.  The source identified as the dominant YSO in 
the NE cluster appears to be older than the dominant
YSO in the cluster associated with the IRAS source, but 
its observed flux seems to be contaminated by 
extra emission, which suggests the presence of a young 
source contributing to the SED at far-IR wavelengths.  The
star formation observed in the field of IRAS~07422-2001 
supports the idea of hierarchical formation of massive 
star clusters and the growth of massive young stellar objects
near the centres of multiple sub-clusters in a star forming 
clump through competitive accretion.
}


\maketitle
%

\section{Introduction}

Studies of massive star formation have been getting significant
attention in the recent years.  The intriguing problem is to 
understand how massive stars form - whether it is through a
process similar to that for low mass stars, ie., through disk 
accretion and driving molecular outflows (e.g. Yorke \& Sonnhalter 
\cite{yorke02}) or through alternative scenarios like coalescence
of lower mass stars (e.g. Bonnell, Bate \& Zinnecker \cite{bonnell98}). 
Even though it is still debatable how very massive stars form, it is 
almost clear that stars at least up to late-O spectral types form 
primarily through disk accretion and that the poor collimation
of outflows from many of these objects previously seen in
single-dish CO observations is mainly due to multiple 
collimated outflows (eg. Beuther et al. \cite{beuther02}, 
Varricatt et al. \cite{varricatt10}).
However, it has been proposed that their formation may be 
different from low mass stars' in the sense that they may form 
mainly in massive clusters
through competitive accretion, where the cluster potential aids
the growth of massive YSOs near the centres of massive clusters
(eg. Bonnell \& Bate \cite{bonnell06})

IRAS~07422-2001 (henceforth IRAS~07422) is a luminous YSO, first 
detected in $^{12}$CO(J=1-0) by Wouterloot \& Brand (\cite{wb89})
using the 15-m SEST.  They detected a blue wing emission in CO, 
implying the presence of outflows.
From a  V$_{lsr}$=52.7\,km~s$^{-1}$  of the CO 
line, they derived a kinematic distance of 5.1\,kpc 
and a luminosity of 7.4\,$\times$10$^3$\,L$_{\odot}$. 
Palla et al. (\cite{palla93}) detected strong H$_2$O masers at
v$_{peak}$=44.8\,km~s$^{-1}$ and derived a kinematic distance
of 4.2\,kpc.  Brand et al. (\cite{brand94}) also
detected H$_2$O masers from this source at similar radial velocity
and flux density.

Slysh et al. (\cite{slysh99}) did not find any 6.7\,GHz
class-II methanol maser emission from this region
at a 3$\sigma$ sensitivity of 3\,Jy suggesting that the YSOs here 
may be very young.  Codella \& Felli (\cite{codella95}) did 
not detect any H89$\alpha$ hydrogen recombination line emission 
from IRAS~07422 suggesting that this source is very young and 
has not yet formed a diffuse H{\sc{ii}} region.
Urquhart et al. (\cite{urquhart07}), in their Red MSX Source (RMS) 
radio survey, did not detect any emission
(above their rms noise levels of 0.14\,mJy and 0.17\,mJy at 3.6\,cm
and 6\,cm respectively) from IRAS~07422 
indicating that the YSOs detected here at mid- and far-IR wavelengths
are in pre-UCH{\sc{ii}} phase.

A dense core was detected towards this source in 97.981-GHz CS(2-1) emission
(Bronfman, Nyman \& May \cite{bronfman96}) at V$_{lsr}$=52.8\,km~s$^{-1}$;
Slysh et al. (\cite{slysh97}) observed OH thermal emission
at a similar velocity (52.3\,km~s$^{-1}$).
These are significantly different from the velocity at 
which H$_2$O masers were detected.  However,  H$_2$O masers seen
towards massive YSOs are understood to be excited by jets
(eg. Felli, Palagi, Tofani \cite{felli92}, Goddi et al. \cite{goddi05}).
Hence we use the 52.8 km~s$^{-1}$ of the CS emission for estimating
the distance.  Using the latest rotation curves of Reid et al. 
(\cite{reid09}), we estimate a kinematic distance of 
4.01\,(+0.69, -0.64)\,kpc towards IRAS~07422.  We will use this 
distance for the calculations in this paper.  

Even though outflow signatures were observed in the data of 
Wouterloot \& Brand (\cite{wb89}), no systematic observations have 
been reported on this source to investigate it further.  Near-IR 
imaging is a powerful tool to study star forming regions.
Outflows from YSOs emit strongly in the near-IR ro-vibrational
lines of H$_2$;  the H$_2$ $v$=1-0 S(1) line at 2.1218\,$\mu$m
is a powerful tracer of outflows via
shock excitation.  The high spatial resolution available with
moderately sized telescopes at IR wavelengths when compared to
that available with single-dish radio and mm telescopes
and the much lower interstellar extinction at IR wavelengths
compared to that at optical wavelengths make this regime ideal
for studying massive star forming regions.
In this paper, we use near-IR imaging to understand 
the star formation in IRAS~07422.  The new data are used along 
with available archival data to study the properties of the YSOs.


\section{Observations and data reduction}

\subsection{WFCAM imaging}
\label{ukirtdata}
Observations were obtained with the 3.8-m United Kingdom 
Infrared Telescope (UKIRT) and the Wide Field Camera (WFCAM{\footnote{http://www.jach.hawaii.edu/UKIRT/instruments/wfcam/}; 
Casali et al. \cite{casali07}) using near-IR broadband $J$, 
$H$ and $K$ MKO filters and a narrow-band MKO filter centred 
at the wavelength of the H$_2$ (1-0) S1 line at 2.1218\,$\mu$m. 
WFCAM has a pixel scale of 0.4$\arcsec$~pix$^{-1}$ and employs 
four 2048$\times$2048 HgCdTe HAWAII-2RG arrays, each with
with a field of view of 13.65$\arcmin\times$13.65$\arcmin$.
The four arrays are arranged in a square pattern in the focal
plane, with a gap of 12.83$\arcmin$ each in the fields covered by 
adjacent arrays.  For the current work, we used the data from 
only one of the four arrays in which the object was located.
Observations were performed by dithering the object (to 9 points 
in the focal plane for $J, H$ and $K$ bands and to 5 
points for H$_2$) separated by a few arcseconds.
We used a 2$\times$2 microstepping, hence the final pixel scale 
is 0.2$\arcsec$~pixel$^{-1}$.  The spatial resolution is limited 
by the seeing at the time of observations.  

The first set of observations were obtained on 20110218~UT.
Since outflows were detected in the H$_2$ line, we further
observed the region in H$_2$ and $K$ filters on three more 
epochs to improve the depth of detection.
Table \ref{tab:obslog} gives a log of the WFCAM observations
performed.

\begin{table}
\caption{Log of WFCAM observations}     
\label{tab:obslog}      		
\centering                      	
\begin{tabular}{llllll}        		
\hline\hline                                    \\[-2mm]
UTDate 		&Filters	&Exp. time &Total int.		&FWHM		\\[1mm]
({\small{yyyymmdd}})&used		&(s)	&time (s)	&(arcsec) 	\\
\hline	\\[-2mm]
20110218	&$J, H,$	&10, 5, &720, 360, 		&0.83, 0.81, \\[1mm]
		&$K, H_2$ 	&5, 40  &360, 800     		&0.82, 0.81  \\[1mm]
20110309	&$K, H_2$	&5, 40	&360, 800		&1.40, 1.28  \\[1mm]
20110310	&$H_2$      	&40	&800			&1.16	     \\[1mm]
20110311        &$H_2$      	&40	&800			&1.37	     \\[1mm]
\hline
\end{tabular}
\end{table}

The data were reduced by the Cambridge Astronomical Survey Unit 
(CASU); the archival and distribution of the data are carried 
out by the Wide Field Astronomy Unit (WFAU).  The sky conditions 
were not photometric during the observations.  However, since 
the flux calibration was performed by deriving the photometry 
of isolated point sources with 2MASS detection in the large 
field of each array, the photometric accuracy is good for the 
stars detected.  The derived magnitudes are in the MKO 
photometric system. 
Fig. \ref{wfcamJHH25} shows a $JHH_2$ colour composite image
of a  5.5$\arcmin\times$5.5$\arcmin$ section of the full 
WFCAM image, showing the area where the outflow and star formation 
activity discussed in this paper take place.

\begin{figure*}
\centering
\includegraphics[width=18.2cm]{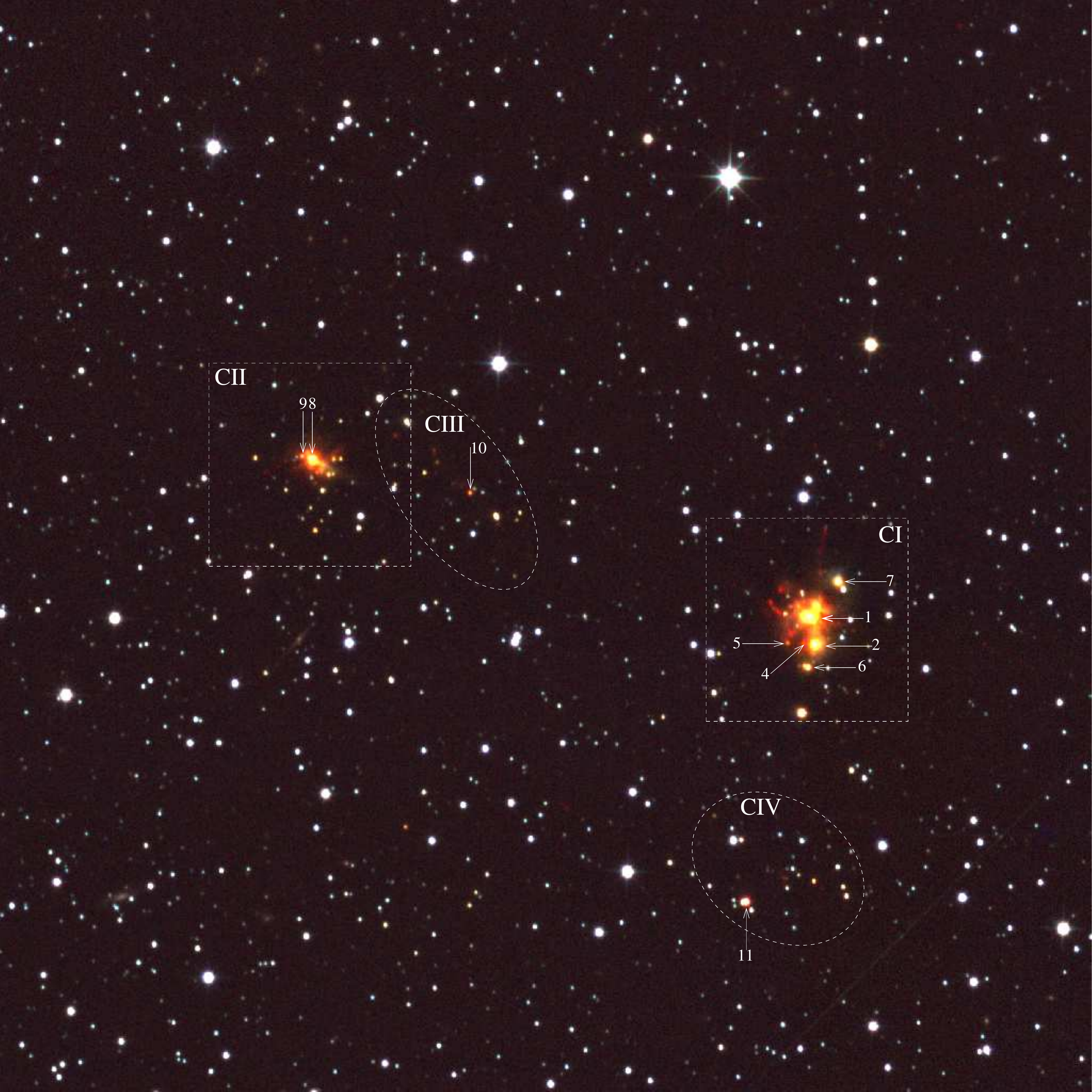}
\caption{WFCAM $JHH_2$ colour-composite image
($J$-blue, $H$-green, $H_2$-red) in a 
5.5$\arcmin\times$5.5$\arcmin$ field covering the
embedded clusters associated with IRAS~17527.}
\label{wfcamJHH25}%
\end{figure*}

The sky background was fitted and removed from the mosaics.
The H$_2$ images were continuum-subtracted using the $K$-band 
images.  The average of the ratio of counts, for a few isolated 
point sources, between the $K$ and H$_2$ images was obtained. 
The background-subtracted $K$-band image was scaled by this ratio 
and was then subtracted from the background-subtracted H$_2$
image.   The H$_2$ mosaic of 20110218 was continuum-subtracted
using an average of the two $K$-band mosaics obtained on the 
same night; those of 20110309--20110311 were continuum-subtracted
using the average of the two $K$-band mosaics obtained on 
20110309. Due to differences in seeing between the $K$ and H$_2$ 
images, point sources give positive and negative residuals.  
A uniform continuum-subtraction over all objects in the field
is difficult to achieve in regions with a wide range in 
near-IR colours.  The $K$/H$_2$ ratio has a dependence on the 
apparent near-IR colours; hence highly reddened sources and 
objects with excess appear with large negative residuals in 
the continuum-subtracted H$_2$ image  
(see Varricatt \cite{varricatt11} for a discussion).
Fig. \ref{wfcamH2} shows the continuum-subtracted H$_2$ image
of the region shown in Fig. \ref{wfcamJHH25}.  Images from all
four epochs of our H$_2$ observations are averaged here to
improve the S/N of the emission features detected. 

\begin{figure*}
\centering
\includegraphics[width=15cm]{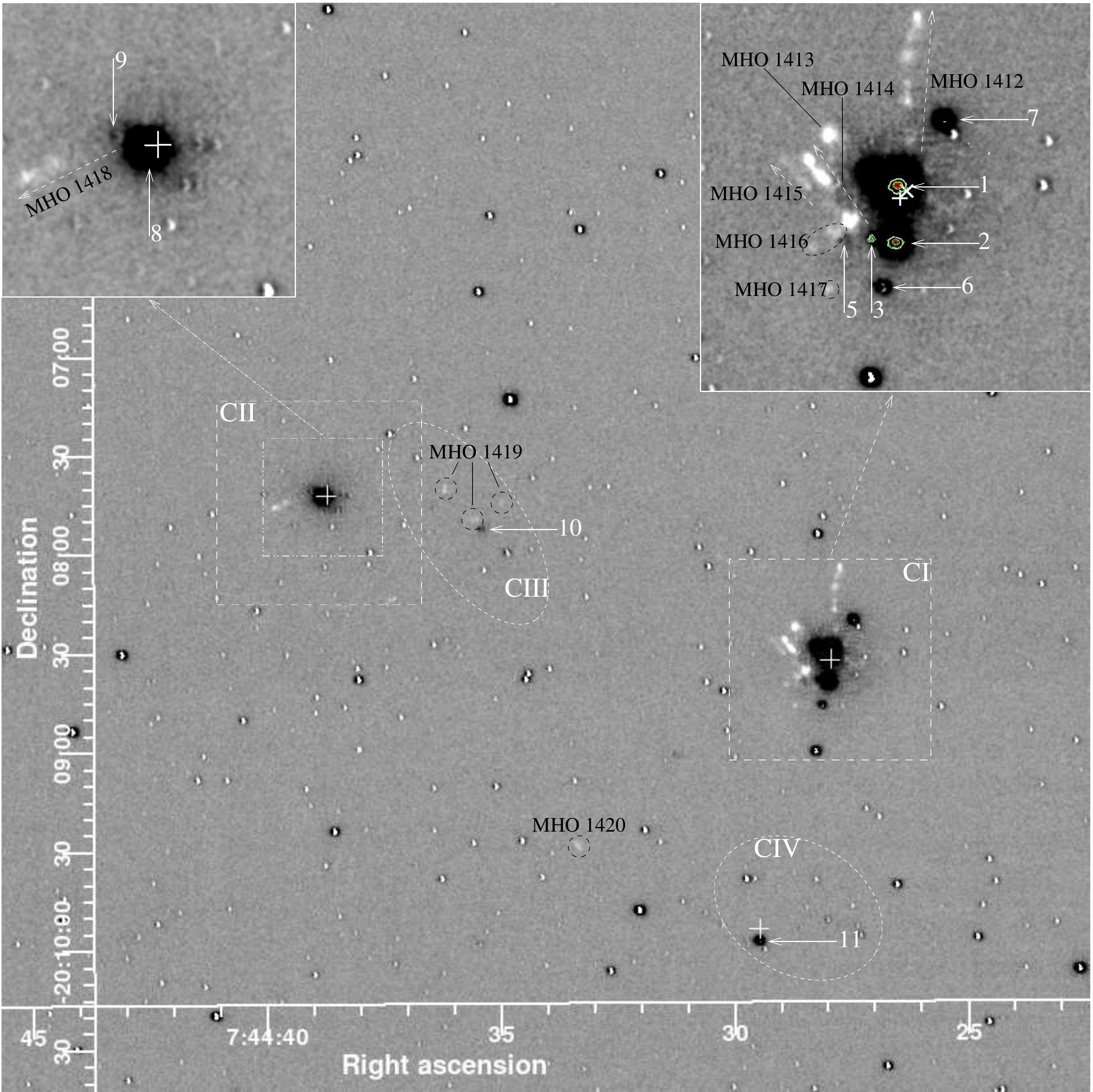}
\caption{The continuum-subtracted 2.122-$\mu$m
image of IRAS~07422 over a 5.5$\arcmin$$\times$5.5$\arcmin$ 
field shown in Fig. \ref{wfcamJHH25}, smoothed with a 
2-pixel FWHM Gaussian to enhance the appearance of the 
faint emission features. ``$\times$'' and ``+'' show the
locations of the IRAS and MSX sources respectively.
The top-right inset shows an expanded view of a 
1$\arcmin\times$1$\arcmin$ field containing ``CI''; the 
contours generated  from the 11.2-$\mu$m Michelle, showing
the 3 objects detected at 11.2$\mu$m (\#1, \#2 and \#3), are overlaid on it.
The top-left inset shows a 35$\arcsec\times$35$\arcsec$ 
field in ``CII''.
}
\label{wfcamH2}
\end{figure*}

\subsection {Archival data}

The IRAS mission detected a point source at $\alpha$=7:44:27.9 
$\delta$=-20:08:32 (J2000) in all four bands (at 12, 25, 60 
and 100\,$\mu$m). IRAS~07422 was detected well in the all 
sky survey conducted by the AKARI satellite 
(Murakami et al., \cite{murakami07}).  The 
mid-IR camera - Infrared Camera (IRC) - on board AKARI 
detected three sources at 18\,$\mu$m.  The brightest two
of these were also detected by the AKARI Far-Infrared 
Surveyor (FIS) in all 4 bands (at 65, 90, 140 and 160\,$\mu$m).  

At mid-IR wavelengths, the MSX mission (in four bands 
A, C, D and E centred at 8.28, 12.13, 14.65 and 
21.23\,$\mu$m respectively) also detected three sources 
in this region. The locations of the MSX sources roughly 
coincide with those of the three sources detected by 
AKARI-IRC at 18\,$\mu$m.  This region was observed in 
the Wide-field Infrared Survey Explorer 
(WISE; Wright et al. \cite{wright10}) mission in four 
bands - $W$1 (3.4\,$\mu$m), $W$2 (4.6\,$\mu$m), 
$W$3 (12\,$\mu$m) and $W$4 (22\,$\mu$m).  The images in 
the four bands have angular resolution of 6.1, 6.4, 6.5 
and 12.0$\arcsec$ respectively and have an astrometric 
precision better than 0.15\arcsec for sources detected 
with good S/N.  The excellent sensitivity of the WISE data 
enables detection of many faint features in the data.  

Most of the mid-IR observations available on this source 
have poor angular resolution compared to our 
near-IR WFCAM images.  No {\it Spitzer} data is 
available for IRAS~07422.
The only mid-IR observation available
with good angular resolution is with UKIRT and Michelle.  
Michelle (Glasse et al. \cite{glasse97}) is a mid-infrared 
imager/spectrometer operating in the 8--25\,$\mu$m wavelength 
regime, using an SBRC Si:As 320x240-pixel array.  It has 
an image scale of 0.21\arcsec/pix and a field of view of 
67.2\arcsec$\times$50.5\arcsec at the Cassegrain focal 
plain of UKIRT.  The Michelle observations were performed 
on 20040422 (UT) using an $N'$ filter centred at 11.2\,$\mu$m 

The Michelle data were downloaded from UKIRT data archive and 
were reduced using the facility data reduction pipeline ORACDR.
The observations were performed by chopping the secondary and 
nodding the telescope by 20\arcsec each (peak-peak) in
mutually perpendicular directions to detect faint sources in 
the presence of strong background emission at thermal 
wavelengths.  The resulting mosaic has four images of the 
source (2 positive and 2 negative beams). The final image was 
constructed after negating the negative beams and combining
all 4 beams.  Fig. \ref{mich11p2} shows the final mosaic of 
the region around IRAS~07422.  An average FWHM of 0.8\arcsec
was measured from the sources detected in the 11.2\,$\mu$m
image.


\section{Results and discussion}

Our near-IR images reveal a very active picture of star
formation associated with IRAS~07422.  Fig. \ref{wfcamJHH25} 
shows a large number of red sources in at least two
embedded clusters and two additional cluster candidates
active in star formation.  Regions containing them
are enclosed in dashed boxes or 
ellipses in the figure and are labelled ``CI--CIV''; the 
most prominent of these are ``CI'' and ``CII''.  Interesting 
point sources are labelled ``1--11'' on the figure.

The continuum-subtracted H$_2$ image (Fig. \ref{wfcamH2})
shows multiple outflows from this region.  Dashed arrows 
are shown on Fig. \ref{wfcamH2} where aligned H$_2$ line 
emission knots are present, which suggest the directions 
of the outflows.  The H$_2$ emission features are 
named MHO~1412--1420 in the Catalogue of Molecular
Hydrogen emission-line Objects{\footnote
{http://www.jach.hawaii.edu/UKIRT/MHCat/}} (Davis et al.
\cite{davis10}) and are labelled on the figure.
Table \ref{tab:outflow} shows the directions
and lengths of the outflows detected in Fig. \ref{wfcamH2}.
The angles given in the table are east of north,
measured from our H$_2$ image.

Near-IR ($J-H$) - ($H-K$) colour-colour diagram is a powerful
tool to identify YSO candidates via reddening and excess.
The $JHK$ magnitudes of point sources detected over the 
13.5\arcmin$\times$13.5\arcmin field covered by the array 
on which IRAS~07422 was located are used to construct the 
colour-colour diagram. Magnitudes of the sources detected 
over multiple epochs are averaged. We had clouds present 
during the observations.  Therefore, to avoid any influence 
from variable clouds, objects located within 7\arcsec of 
the edges of our near-IR images are omitted from the 
colour-colour diagram so that only those present in all 
jittered frames are used. Only those objects with a 
photometric error of 0.207\,mag or less in $K$, which is 
our 1-$\sigma$ error in $K$ at 19\,mag, are used.  
Fig. \ref{07422JHKcol}
shows the colour-colour diagram constructed from our
$JHK$ observations.  The continuous and the short dashed green
curves show the loci of the colours of main-sequence and 
giant stars adopted from Tokunaga (\cite{tokunaga00}).  
The long dashed green line shows the loci of Classical T
Tauri stars (CTTS) from Meyer, Calvet \& Hillenbrand (\cite{meyer97}).
The dotted red arrows show the reddening vectors 
up to A$_V$=50 derived from the interstellar extinction law
given in Rieke \& Lebofsky (\cite{rieke85}).  The objects located
in between the two main-sequence reddening vectors are expected
to be reddened main-sequence stars.  
YSOs occupy the region of the colour-colour diagram
below the lower reddening vector for the main-sequence stars, where
CTTS, Herbig Ae/Be (HAeBe) stars and Luminous YSOs occupy different
regimes (Lada \& Adams \cite{lada92}).  

\begin{figure*}
\centering
\includegraphics[width=15cm]{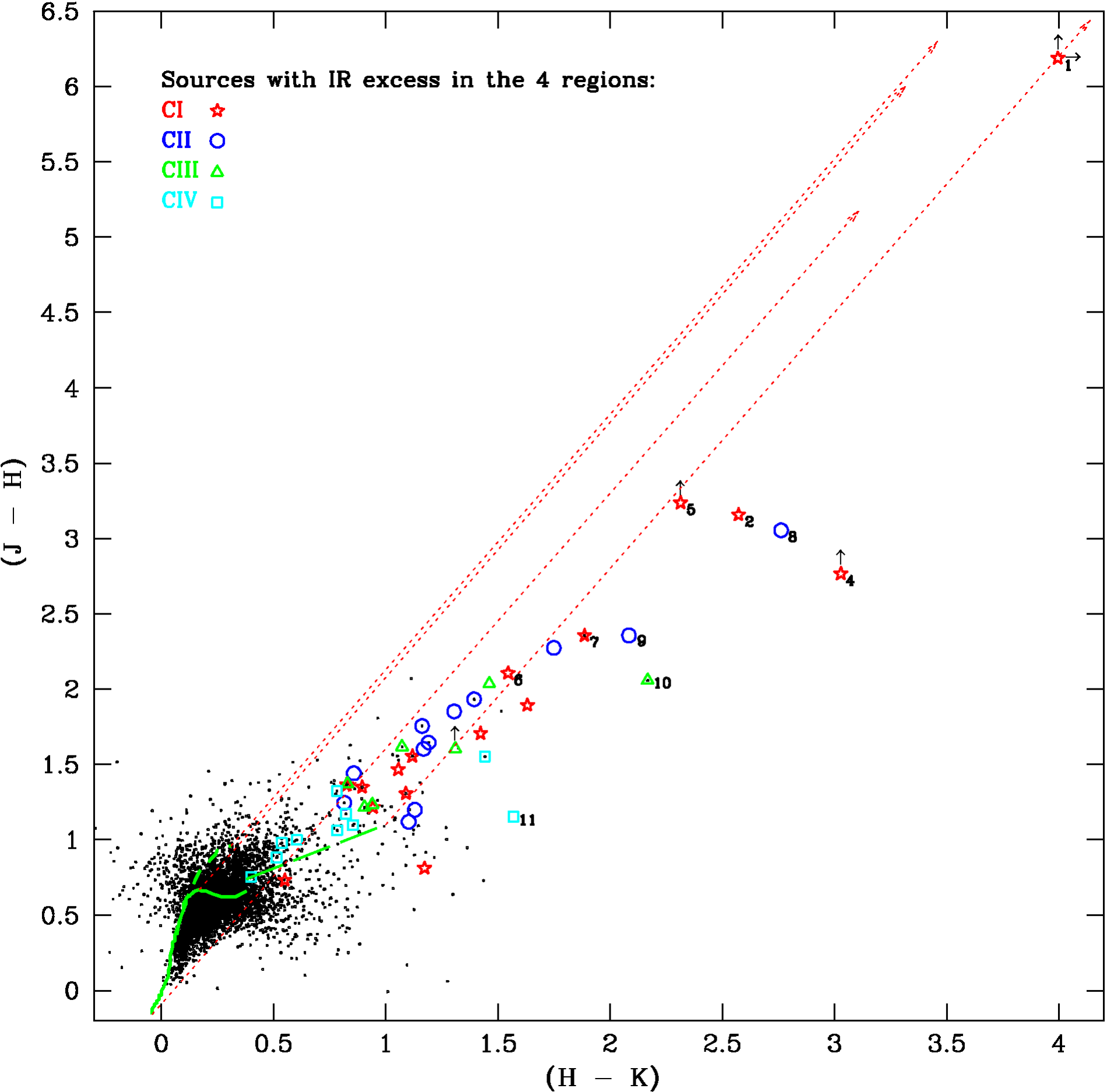}
\caption{The $JHK$ colour-colour diagram of the
13.5\,$\arcmin$ $\times$ 13.5\,$\arcmin$ field surrounding IRAS~07422.
The curved lines show the loci of main sequence (continuous, green)
and giant (short-dashed, green) stars from Tokunaga(\cite{tokunaga00}).
The long-dashed green line shows the loci of CTTS from Meyer et al. (\cite{meyer97}).
The dotted arrows (red) show the reddening vectors for A$_V~\leq$~50
derived from the interstellar extinction law given by Rieke \& Lebofsky (\cite{rieke85}).
The black dots show objects with photometric error less than 0.207 mag, which
is our 1-$\sigma$ error for a source of 19\,mag in  $K$.  Sources which are fainter
than 19 mag in $K$ are not shown here. Objects with upper limit magnitudes
in $J$ are shows with an upward directed arrow; source \#1 with $H$ magnitude also is treated
as an upper limit is shown with upward and rightward directed arrows.}
\label{07422JHKcol}
\end{figure*}

Fig. \ref{wfcamK} shows the WFCAM $K$-band image of the
5.5$\arcmin\times$5.5$\arcmin$ field shown in Fig. \ref{wfcamJHH25}.
All sources in this field exhibiting IR excess in the $JHK$ colour-colour
diagram (Fig. \ref{07422JHKcol}) are enclosed in green circles; 
sources with IR excess in regions ``CI--CIV'' are enclosed in
red boxes.   Contours are overlaid on Fig. \ref{wfcamK} to show
the number of objects with IR excess, within a radius of
10\arcsec, estimated at 1\arcsec intervals over the 
5.5$\arcmin\times$5.5$\arcmin$ field.  The contours
show enhancements in the number density of sources with IR excess
towards regions ``CI--CIV'' suggesting 
that these regions host embedded clusters.

\begin{figure*}
\centering
\includegraphics[width=15cm]{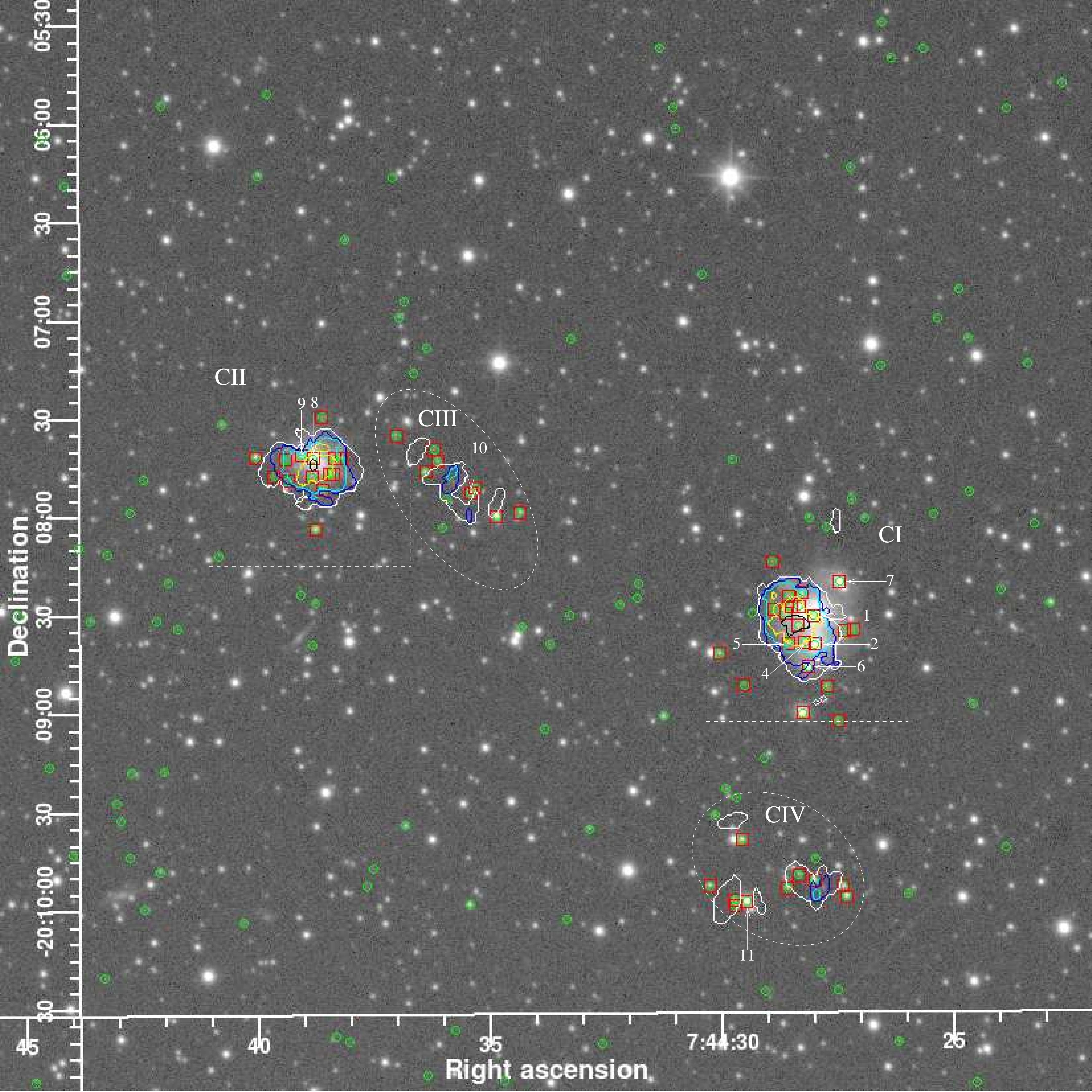}
\caption {WFCAM $K$-band image of the
5.5$\arcmin\times$5.5$\arcmin$ field shown in Fig. \ref{wfcamJHH25}.
All objects in this region exhibiting IR excess in the $JHK$ colour-colour
diagram (Fig. \ref{07422JHKcol}) are enclosed in green circles.
The sources in regions ``CI--CIV'' are enclosed in
red boxes.  The contours in black, yellow, cyan, blue and white respectively
enclose regions with 10, 8, 6, 5 and 4 sources with IR excess in a circle
of 10\arcsec radius, estimated at intervals of 1\arcsec.}
\label{wfcamK}%
\end{figure*}

\begin{figure}
\centering
\includegraphics[width=8.9cm]{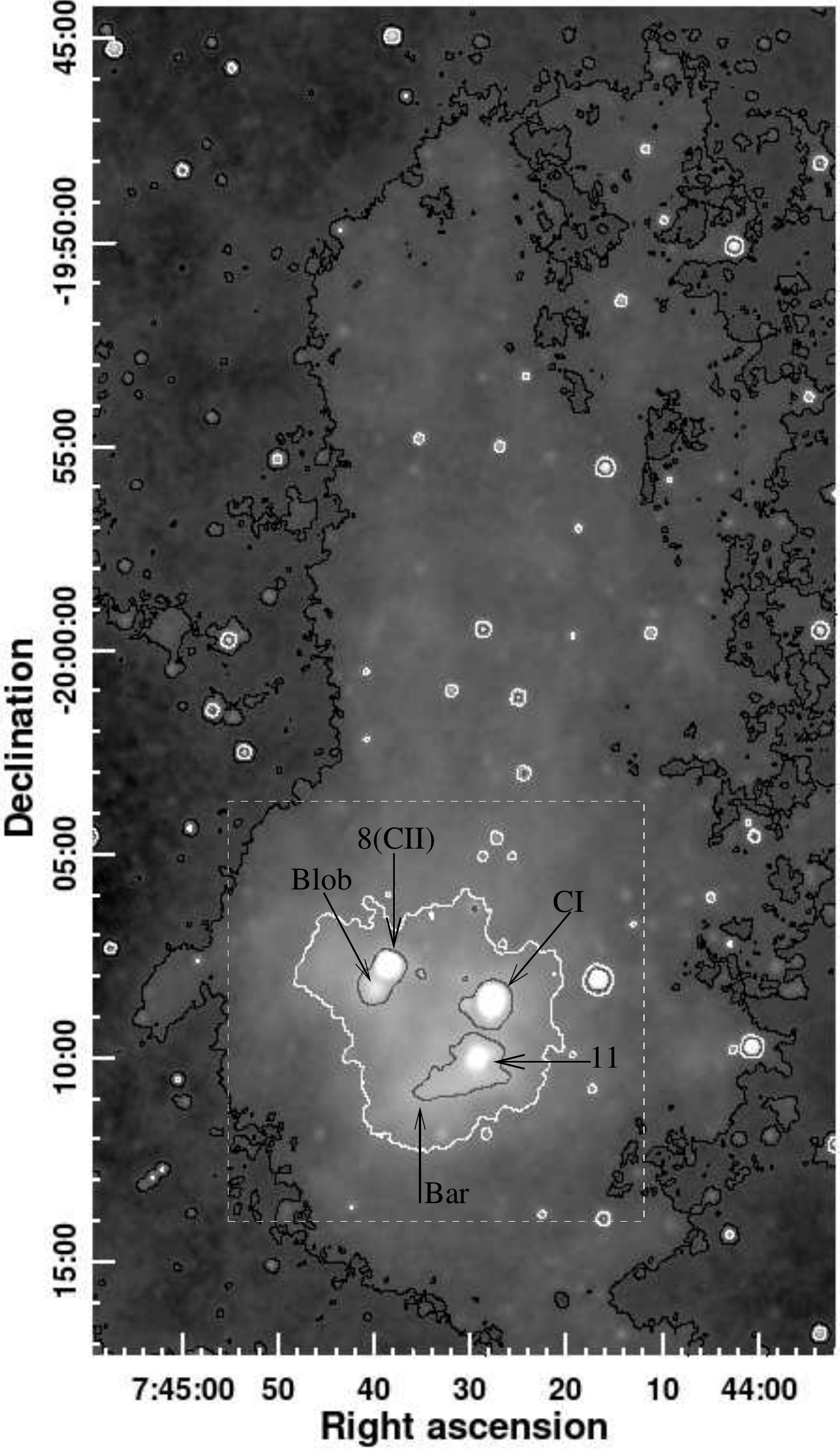}
\caption{The WISE 12-$\mu$m image over a field of
18$\arcmin\times$33$\arcmin$ encompassing the cloud in which IRAS~07422
is located.  The contours in black, white and grey show the boundaries
where the emission from the cloud falls to 15\%, 40\% and 85\% respectively
of the peak flux in the bar. Most of the star formation happening within
this cloud is in the region within the white contours.}
\label{wise3}%
\end{figure}

\begin{figure}
\centering
\includegraphics[width=8.9cm]{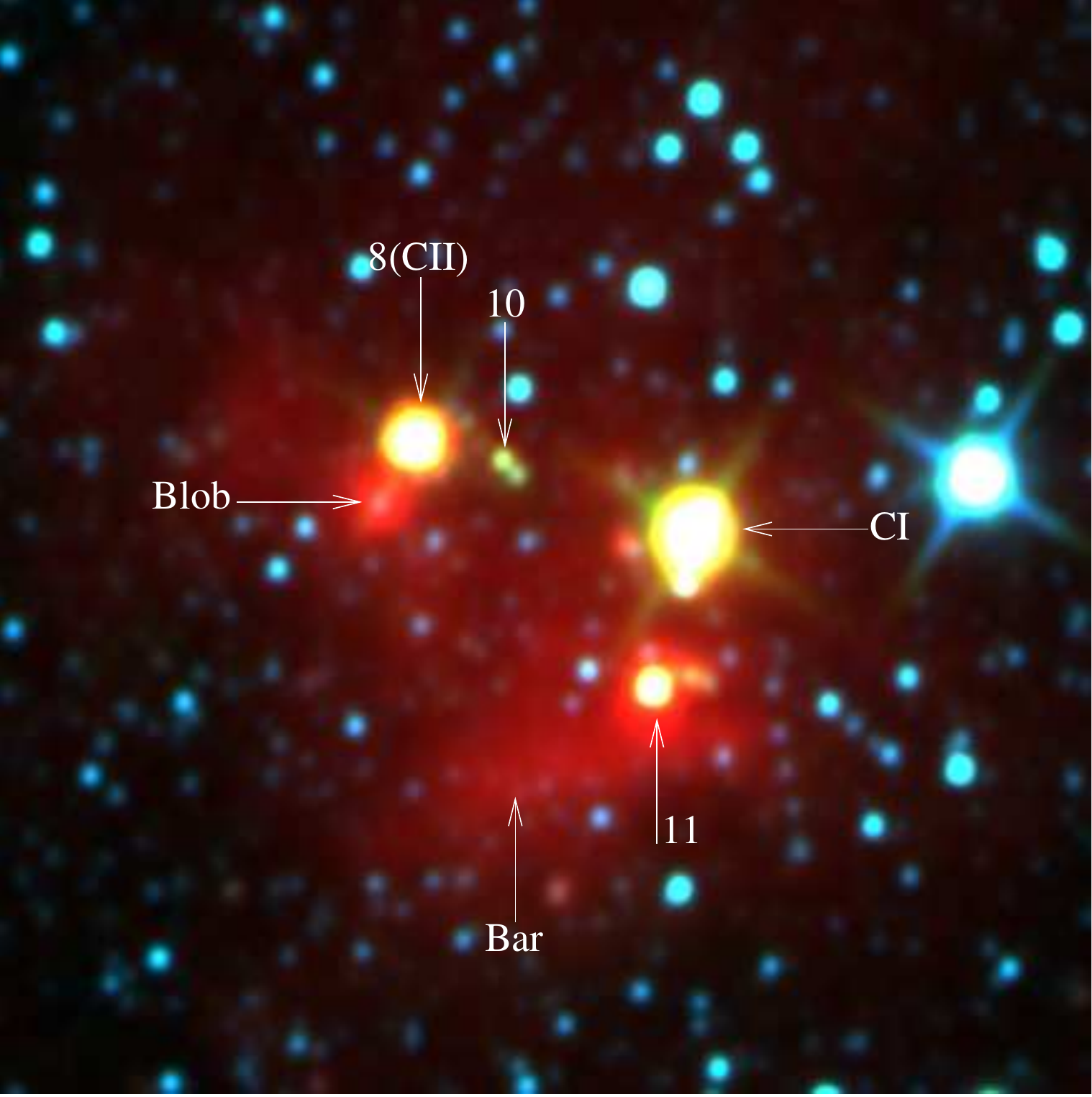}
\caption{The WISE colour-composite image
(3.4\,$\mu$m-blue, 4.6\,$\mu$m-green, 12\,$\mu$m-red) in a
10$\arcmin\times$10$\arcmin$ field (the region enclosed
in box in Fig. \ref{wise3}) including
the embedded clusters and the bright region of the cloud
containing them.}
\label{wise123}%
\end{figure}

The WISE 3.4--22-$\mu$m images provide a deeper view of the star
forming activity in this region. The 12-$\mu$m image (Fig. \ref{wise3}) 
shows a cloud of $\sim$15\arcmin$\times$30\arcmin extending NS.  
The 22-$\mu$m image also shows the nebulosity from the whole 
cloud, albeit with lower S/N than in the 12-$\mu$m image. 
The cloud appears cometary; most of the star formation 
activity seems to be taking place in the ``head'' - in a
region of $\sim$6$\arcmin$ diameter located towards the southern
region of the cloud (region encompassed by the white contour in 
Fig. \ref{wise3}).  This region of the cloud exhibits extended 
nebulosity in all four bands and hosts all the young clusters 
discussed here.  Fig. \ref{wise123} shows a WISE 3.4, 4.6, 
12\,$\mu$m colour composite image of a 
$\sim$10\arcmin$\times$10\arcmin field around the southern 
region of the cloud.  The nebulosity, which is much
stronger at 12\,$\mu$m than at shorter WISE wavelengths, is 
seen red in colour.  

The three MSX sources detected near IRAS~07422 are located in
``CI'', ``CII'' and ``CIV'' (close to the near-IR sources \#1,
\#8 and \#11 respectively).  The three sources detected by 
AKARI-IRC at 18\,$\mu$m agree with the MSX sources within 
positional errors.  The AKARI-FIS detected emission from 
``CI'' and ``CII'' in all four bands.  The FIS source
associated with ``CI'' is the brighter one.  

The outflows detected in H$_2$ and the interesting near-IR 
sources detected are described throughout the text.  The 
coordinates and magnitudes of the prominent sources are 
given in Table \ref{tab:srcflx}.

\subsection{The outflows detected and their possible driving sources}
\label{outflow}

\subsubsection{CI}
\label{CI}

``CI'' is the most prominent feature in the near-IR images
(Fig. \ref{wfcamJHH25}).  This region hosts a cluster of
red sources embedded in nebulosity. An expanded view of 
a 1\arcmin$\times$1\arcmin field containing ``CI'' is shown
in Fig. \ref{wfcamJHH21}.   A few point sources of interest
are labelled ``1--7'' on the figure.  The brightest of
these are the two red near-IR sources labelled  \#1 and \#2.  
In the WFCAM images, \#1 appears to be located behind
an extinction lane running NE-SW. Strong nebulosity is seen 
in the vicinity of \#1; the nebula is roughly bipolar in
appearance in $J$ and $H$ bands with the two blobs of
diffuse emission located NW and SE of the location of
\#1 and of the extinction lane.
$J$, $H$ and $K$ images of the central region
of ``CI'' containing \#1 and \#2 are shown separately as
insets on Fig. \ref{wfcamJHH21}.

\begin{figure}
\centering
\includegraphics[width=8.9cm]{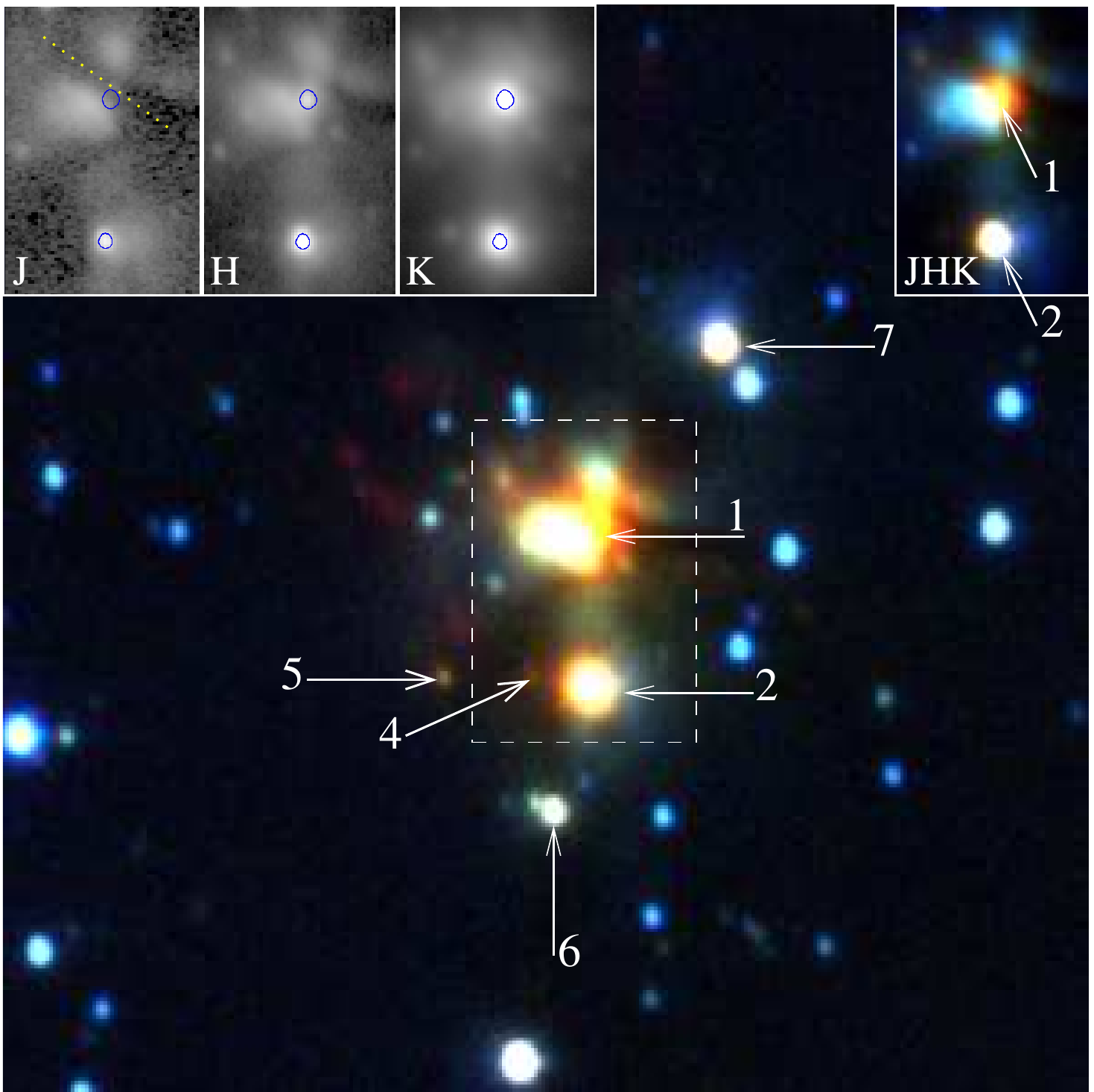}
\caption{WFCAM $JHH_2$ colour-composite image
($J$-blue, $H$-green, $H_2$-red) in a
1$\arcmin\times$1$\arcmin$ field containing ``CI''.
The top-left inset shows $J$, $H$ and $K$ images of the
central 12\arcsec$\times$18\arcsec field (enclosed in the
dashed box) at a higher contrast, with circles showing
the locations of the two bright sources derived
from the $K$-band image;  a dotted yellow line is 
drawn on the $J$-band image along the extinciton lane 
to guide the eyes. The top-right inset shows a $JHK$ colour
composite of the same region ($J$-blue, $H$-green, $K$-red), 
revealing the bright point sources embedded in the nebulosity.}
\label{wfcamJHH21}%
\end{figure}

\begin{figure}
\centering
\includegraphics[width=8.9cm]{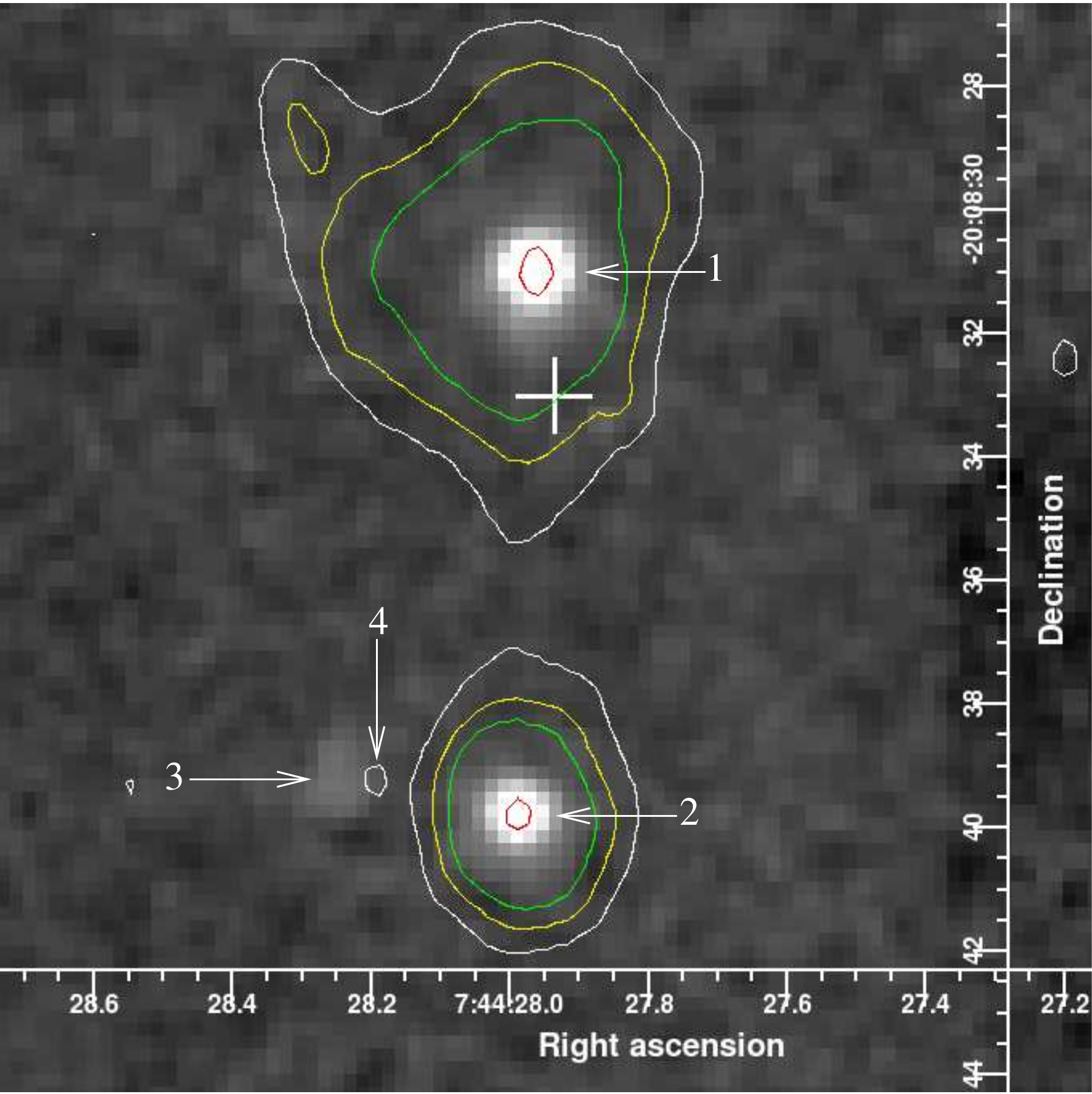}
\caption{Michelle image of IRAS~07422 at 11.2\,$\mu$m
covering a field of 18$\arcsec\times$18$\arcsec$.
Contours from our WFCAM $K$-band image are overlaid.
The ``+'' shows the location of the MSX source.}
\label{mich11p2}%
\end{figure}

The point source \#1 
is bright in $K$ (10.52 mag) and is much fainter in 
$H$ (14.51 mag). In $H$ band, the emission from the SE nebula 
is stronger than that from the point source detected.  In $J$,
\#1 is not detected and only the nebulosity associated 
with it is seen.  Since \#1 is closely associated with 
nebulosity, we derived its $H$ and $K$ magnitudes by performing 
aperture photometry in a 1.5\arcsec-diameter aperture.
\#1 is not detected
in $J$, so we have used the $J$ detection limit to derive 
its $J-H$ colour.  The $H$ magnitude derived may still be 
significantly affected by nebulosity. Hence we treat 
the $H$ magnitude also as an upper limit only for 
deriving its colours. Fig. \ref{07422JHKcol} shows 
that \#1 is the reddest object detected in this region. 
\#2 is located $\sim$8.85\arcsec S of \#1.  \#2 shows large 
reddening and excess,
but is much less red than \#1.  There are several 
other sources in ``CI'', which exhibit large reddening 
and excess  in Fig. \ref{07422JHKcol}.  ``4--7'' are the 
prominent ones among them.  Prominent sources in 
``CI'' with IR excess are shown by ``star'' symbols in
Fig. \ref{07422JHKcol}.

The brightest of the three AKARI-IRC sources is located in 
``CI'', between \#1 and \#2;  the brighter of the two AKARI-FIS 
sources is in ``CI''.  ``CI'' also hosts the brightest 
of the three MSX sources, which is detected in all four MSX
bands.  The location of the brightest of the sources 
detected by IRC, FIS and MSX coincides with the IRAS 
source within the positional errors.

\begin{table*}
\caption{A few of the interesting sources detected}
\label{tab:srcflx}      
\centering                      
\begin{tabular}{llllll}        
\hline\hline                                    \\[-2mm]
Source  &Coordinates (J2000)    &$J$    &$H$    &$K$    &11.2-$\mu$m flux        	\\
&($\alpha$, $\delta$)           &(mag)$^{\mathrm{a}}$  &(mag)$^{\mathrm{a}}$  &(mag)$^{\mathrm{a}}$  &(mJy)$^{\mathrm{a}}$                                  		\\
\hline                                                                                                   	\\[-2mm]
1$^{\mathrm{b}}$  &07:44:27.964, -20:08:30.98    &               &14.511 (0.002) &10.516 (0.020) &579 (31)      \\
2$^{\mathrm{b}}$  &07:44:27.988, -20:08:39.77    &16.475 (0.003) &13.317 (0.001) &10.745 (0.020) &364 (25)      \\
3		  &07:44:28.251, -20:08:39.28    &               &               &               &80  (15)      \\
4$^{\mathrm{b}}$  &07:44:28.196, -20:08:39.19    &               &17.933 (0.031) &14.905 (0.010) &              \\
5                 &07:44:28.552, -20:08:39.30    &               &17.463 (0.016) &15.149 (0.013) &              \\
6                 &07:44:28.123, -20:08:46.77    &17.227 (0.030) &15.120 (0.001) &13.575 (0.006) &              \\
7                 &07:44:27.457, -20:08:20.92    &16.361 (0.002) &14.005 (0.004) &12.119 (0.002) &              \\
8$^{\mathrm{b}}$  &07:44:38.814, -20:07:42.27    &16.920 (0.014) &13.865 (0.004) &11.103 (0.002) &              \\
9$^{\mathrm{b}}$  &07:44:39.035, -20:07:41.31    &19.360 (0.051) &17.003 (0.018) &14.919 (0.010) &              \\
10                &07:44:35.411, -20:07:52.88    &19.351 (0.053) &17.293 (0.003) &15.125 (0.002) &              \\
11                &07:44:29.476, -20:09:57.95    &15.425 (0.015) &14.271 (0.001) &12.701 (0.002) &              \\
\hline
\end{tabular}
\begin{list}{}{}
\item[$^{\mathrm{a}}$] The values given in parenthesis are the  1-$\sigma$ errors.
\item[$^{\mathrm{b}}$] The $JHK$ magnitudes for these sources are aperture photometry in a 1.5\arcsec-diameter aperture; there may still be contamination from nebulosity.
\end{list}
\end{table*}

Fig. \ref{mich11p2} shows the 11.2-$\mu$m image obtained 
using UKIRT and Michelle.  An 18\arcsec$\times$18\arcsec
field, which encompasses the objects detected at 11.2-$\mu$m,
is shown in the figure.   Only three objects are detected 
in the Michelle image;  \#1 and \#2 are the brightest.
Astrometric corrections are applied to 
the Michelle image, using the coordinates of \#1 and \#2 
from our WFCAM images.  The third source (labelled ``3'') 
is very faint; it is located 3.8\,arcsec NE of \#2.  
The coordinates of \#1 and \#2 from the WFCAM images and 
that of \#3 measured from the calibrated Michelle image 
are listed in Table \ref{tab:srcflx}.  The 11.2-$\mu$m 
flux obtained from the Michelle observations are also 
listed in the table.   \#1 is the brightest source
in the Michelle image at 11.2\,$\mu$m and \#3 is the 
faintest.  \#3 is detected above 5$\sigma$ at 11.2 micron.
We do not see any $JHK$ counterpart
at exactly the same location as of \#3.  However,  
Fig. \ref{wfcamJHH21} shows a source labelled \#4 located
only 0.72\arcsec W of \#3.  This offset is significant, considering
the positional accuracy of our WFCAM observations. Contours
generated from our WFCAM $K$-band image are overlaid on the
Michelle image in Fig. \ref{mich11p2}, which show the 
detection of  \#1 and \#2 at 11.2\,$\mu$m and the offset
of \#3 from \#4.  However, since the Michelle image is 
astrometry calibrated with the coordinates of only two 
sources from WFCAM (\#1 and \#2), we need more
observations, especially at thermal IR to ascertain if  \#3
and \#4 are the same or different.  \#4 is a highly reddened
source.  It is detected well in $K$, marginally detected 
in $H$ and is not detected in $J$.

Our continuum-subtracted H$_2$ image (Fig. \ref{wfcamH2}) 
reveals several line emission features in the vicinity of the 
cluster. These are labelled MHO~1412--1417 on the figure.  The 
possible directions and lengths of some of the outflows implied 
by these line emission knots are shown by dashed arrows on the 
figure and are listed in Table \ref{tab:outflow}.  The most 
prominent of these is MHO~1412, which appears to be directed 
away from \#1,  with H$_2$ knots extending up to a projected
length of ~27\arcsec (0.53\,pc) from \#1.
With \#1 being the reddest and the
brightest source seen in the near- and mid-IR,
and being the primary contributor to the emission 
seen at longer wavelengths, we identify \#1 as the most luminous 
YSO in ``CI''.  From the locations of the features MHO~1412--1417, it 
appears that at least 6 different outflows may be responsible 
for them.  Other than \#1, we identify \#2 and \#3 as two 
other YSOs responsible for
driving the outflows in this region.  However it is difficult to
assign any of the observed H$_2$ line emission features MHO~1413--1417
as originating from these sources.  Some of the other reddened
IR-excess sources in ``CI'' may also be driving outflows,
some of which are seen in the H$_2$ line emission. 
The spectral energy distribution of \#1 and a model fit
to it are discussed in $\S$\ref{sed_1}.

The extinction lane close to source \#1 deserves detailed
investigation.  In $J$ and $H$ bands, it has the appearance
of a nearly edge-on disk at an angle of $\sim$53$^{\circ}$ 
east of north (see the inset in Fig.  \ref{wfcamJHH21}).  
It is also noteworthy that the two H$_2$ knots in MHO~1416 
and the bright knot at the base of MHO~1414 align well at an 
angle of 127$^{\circ}$ east of north (see Fig. \ref{wfcamH2}),  which is roughly 
perpendicular to the orientation of the extinction lane.  
High angular resolution observations at longer wavelengths,
especially in the mm, are required to understand if the 
extinction lane is caused by a nearly edge-on circumstellar 
disk.

\begin{table}
\caption{Outflow lengths and directions}     
\label{tab:outflow}          
\centering                   
\begin{tabular}{llll}       
\hline\hline                                    	\\[-2mm]
Outflow 	&Angle ($^{\circ}$)	&\multicolumn{2}{c}{Length}     \\
ID		&(E of N)		&arcsec 	&parsec  	\\[2mm]
\hline							\\[-2mm]
MHO 1412	&353    &27.2           &0.53      	\\[1mm]
MHO 1414	&38     &$>$14.5        &$>$0.28   	\\[1mm]
MHO 1415	&41     &$>$15.5        &$>$0.30   	\\[1mm]
MHO 1418	&116    &$>$13          &$>$0.25   	\\[1mm]
\hline
\end{tabular}
\end{table}

\subsubsection{CII}
\label{CII}

``CII'' hosts an embedded cluster  $\sim$2.7\arcmin~NE 
of the IRAS position (Fig. \ref{wfcamJHH25}).  An expanded 
view of ``CII'' is shown in Fig. \ref{wfcamJHH2CII}.  The
WFCAM image shows a large number of objects
surrounding a bright source (labelled \#8) detected in the 
near-IR and exhibiting reddening and excess in the
$JHK$  colour-colour diagram (Fig. \ref{07422JHKcol}). 
\#8 is the brightest near-IR source in ``CII''.
It is less embedded than \#1 and
is detected well in our $J$, $H$ and $K$ images.
There is no IRAS detection
in ``CII''.  The second brightest of the AKARI-IRC and
FIS, and MSX sources is detected in this region, but
they lack the positional accuracy to assign the detection to 
\#8 or any other reddened source in the cluster. 
However, the WISE images with very good positional accuracy 
shows a bright source, the position of which agrees well with 
what is derived from the WFCAM images.  Hence, \#8 is
likely to be the most luminous YSO in the cluster.
In addition to \#8, ``CII'' 
hosts several other reddened sources with IR excess. 
Prominent sources with excess, detected well in our images 
are shown with blue circles in Fig. \ref{07422JHKcol}.

\begin{figure}
\centering
\includegraphics[width=8.9cm]{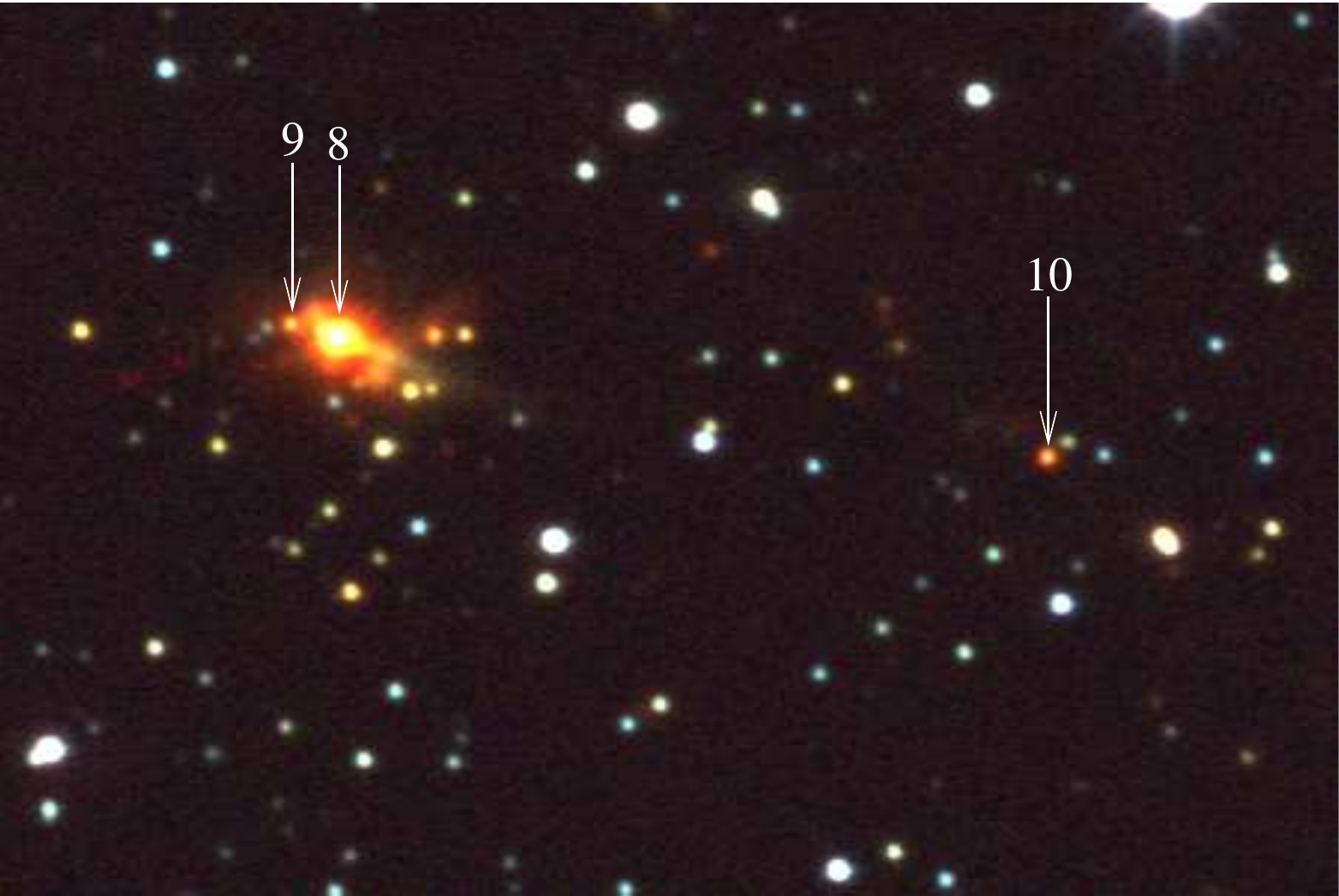}
\caption{WFCAM $JHH_2$ colour-composite image
($J$-blue, $H$-green, $H_2$-red) in a
1.5$\arcmin\times$1.25$\arcmin$ field containing
``CII'' and ``CIII''.}
\label{wfcamJHH2CII}%
\end{figure}

Fig. \ref{wfcamH2} shows aligned H$_2$ line emission knots
labelled MHO~1418 and shown by a dotted arrow.  The direction
of orientation of the H$_2$ knots imply an outflow emanating 
from ``CII''.  It is not clear if \#8 or any of the fainter
reddened sources with excess nearby is responsible for driving 
MHO~1418. The limited H$_2$ knots in MHO~1418 do not trace directly 
to \#8, but they trace back to a source \#9 located 
3.25\arcsec NE of \#8. 
MHO~1418 has a length of 16\arcsec (0.31\,pc) if it is
from \#8 and 13\arcsec (0.25\,pc) if it is from \#9.
Fig. \ref{07422JHKcol} shows that
\#9 also exhibits large reddening and excess.
\#8 is the dominant source in the WISE bands.
In $\S$\ref{sed_8}, we model the spectral energy
distribution of \#8.

The WISE 12-$\mu$m image shows a nebulous emission feature
at $\alpha$=07:44:40.034, $\delta$=-20:08:15.91, labelled
``Blob'' in Figs. \ref{wise3} and \ref{wise123}.  It is
offset $\sim$39\arcsec SE of \#8.  The WISE 3.4 and 4.6-$\mu$m 
images show a point source very close to this position.  
It is not clear if they are the same as ``Blob''.  The 
22-$\mu$m image has much poor spatial resolution.  It
doesn't resolve a source at the location of ``Blob'',
but only shows a slight elongation of the nebulosity 
associated with \#8 in the in the direction of the ``Blob''.
Therefore it remains to be investigated at better spatial 
resolution and higher sensitivity at longer wavelengths 
to learn if ``Blob'' hosts a source much younger than 
\#8.

\subsubsection{CIII and CIV}

Regions ``CIII'' and ``CIV'' show concentrations of red
sources.   These appear elongated and are much less prominent 
than ``CI'' and ``CII''.  Fig. \ref{wfcamK} shows that
there is enhancement in the number density of sources with
IR excess towards ``CIII'' and ``CIV''.  They are shown 
enclosed within ovals in Fig. \ref{wfcamJHH25}.  Deep thermal 
imaging at high angular resolution is required to understand 
if these two regions host embedded clusters.

``CIII'' is seen elongated in the NE--SW direction and is
located close to, and towards the west of, ``CII''.  Several
sources in ``CIII'' show IR excess, which are shown with
green triangles in Fig. \ref{07422JHKcol}.  There is a very
red source, \#10, located somewhat near the centre of ``CIII'',
and showing large reddening and excess in Fig. \ref{07422JHKcol}.
It is detected well in $H$ and $K$ bands and is marginally 
detected in $J$.  \#10 is detected well in all four 
WISE bands; in the WISE 22-$\mu$m image, it is one of the few 
objects detected.  There are several other reddened sources 
in ``CIII''.  In H$_2$ line, we detect three line emission 
knots, which are labelled MHO~1419 in Fig. \ref{wfcamH2}.
It is not clear if all these features are from YSOs
in ``CIII''.  One of these knots is is located very close to \#10; it may
be associated with \#10.  With the WISE source coinciding 
with \#10, we identify it as the dominant YSO in ``CIII''.

``CIV'' is located towards the south of ``CI'';  it also 
hosts several reddened sources as seen in Figs. \ref{wfcamJHH25}
and \ref{07422JHKcol}.
Sources in ``CIV'' with excess are shown with cyan boxes
in Fig. \ref{07422JHKcol}.  The source labelled \#11 is the 
prominent among them and is detected well in $J$, $H$ and $K$}.  
Other than ``CI'' and ``CII'', 
``CIV'' is the only region in Fig. \ref{wfcamJHH25} detected
by MSX and AKARI-IRC.  Most of the reddened sources in ``CIV'' 
except \#11 appear to be distributed in a filamentary pattern 
oriented NE--SW.  \#11 is located somewhat south of those 
reddened sources.  At the spatial resolution of 
WISE, more than one source may be contributing to the WISE
detection at the location of \#11.  However the location of 
the WISE source shows that \#11 is the dominant contributor 
to the flux measured in the WISE bands.  The faintest of the 
three AKARI-IRC sources coincides with the near-IR source 
\#11.  This object was not detected by AKARI-FIS in the 
far-IR.  This is also the faintest of the three sources 
detected by MSX in this region.  ``CIV'' was detected by 
MSX in band A only.  We detect an H$_2$ knot MHO~1420,  NE 
of ``CIV''.  It is not clear if it is due to outflow 
from any of the sources in ``CIV'' or ``CI'' discussed 
here.  The WISE images exhibit a bar-like feature in
the nebulosity below ``CIV'', which is labelled in
Figs. \ref{wise3} and \ref{wise123}.

\subsection{The spectral energy distribution}
\label{fitting}

Out of the four embedded clusters identified in our
data, ``CI'' and ``CII'' are the prominent ones.  
Near-IR sources \#1 and \#8 appear to be the most 
luminous YSOs in these clusters, as inferred from 
their possible association with longer wavelength 
data.  The SEDs of these sources are modelled
in \S\ref{sed_1} and \S\ref{sed_8}.

The WFCAM, Michelle, MSX, WISE, AKARI and IRAS 
data were used to construct the SED of the sources.  
Colour corrections were applied to AKARI and IRAS 
data using the correction factors given in their 
respective point source catalogues (Kataza et al. 
\cite{kataza10},  Yamamura et al. \cite{yamamura10}, 
Beichman et al. \cite{beichman88}).

We used the SED fitting tool of 
Robitaille et al. (\cite{robitaille07}) available 
on-line to model the SEDs. They use a grid of 2D 
radiative transfer models presented in 
Robitaille et al. (\cite{robitaille06}), developed by
Whitney et al. (\cite{whitney03a}, \cite{whitney03b}, etc.).  The 
grid consists of 20000 YSO models with SEDs covering stellar masses 
of 0.1--50\,M$_{\odot}$ and evolutionary stages from the early infall
stage to the late disk-only stage, each at 10 different viewing 
angles.  The results of the SED modelling are presented
in Table  \ref{tab:results}. Note that the model assumes a single object
whereas multiplicity will be an issue in our regions.
Hence, the actual errors in some of the parameters could be
larger than the values given in Table \ref{tab:results} due 
to possible contributions to the observed fluxes of the dominant 
YSOs from neighbouring sources.

\subsubsection{Source \#1}
\label{sed_1}

At the poor spatial resolution of IRAS, AKARI-FIS and MSX,
the source detected in ``CI'' will contain the emission from 
the whole cluster. However, \#1 is resolved from the other 
sources in ``CI'' in the WISE 3.4 and 4.6-$\mu$m data.
The WISE images and the well-resolved Michelle image, both
with good positional accuracy, show that \#1 is the dominant
contributor at longer wavelengths (see Fig. \ref{mich11p2}).  
\#1 has a steeply rising SED.  In $K$ band, the combined 
emission from \#2 -- \#7, the extended nebulosity and
the rest of the objects in the vicinity dominates over 
the emission from \#1; the relative contribution from \#1 
being only $\sim$21\%.  At 11.2\,$\mu$m, \#1 contributes 
56.6\% of the total flux of \#1, \#2 and \#3 combined.   

For $H$, $K$ and 11.2\,$\mu$m, we used the flux estimated
from the WFCAM and Michelle data (see \S\ref{CI}) for 
constructing the SED.  The WISE photometry of \#1 is quite 
free of emission from the neighbours, especially in W1 and 
W2 bands, where \#1 is resolved from \#2 and the rest of 
the cluster members.  However, there appears to be a small 
contribution from the neighbours in the photometry
listed in the catalogue. Hence, for WISE W1 and W2 bands,
we re-determined the photometry of source \#1 by performing
relative photometry using the magnitudes of 
neighbouring isolated point sources in the WISE images 
and using a  10$\arcsec$-diameter aperture.  In
WISE band W3, \#1 is not resolved from \#2 and the rest
of the members of the cluster due to poor 
spatial resolution.  We therefore estimated the photometry
of the whole cluster using a large aperture and applied
a flux correction factor of 0.57, which is the relative
contribution of \#1 towards the flux of \#1, \#2 and \#3 
combined, derived from the Michelle data at 11.2\,$\mu$m
given in Table \ref{tab:srcflx}.
The IRAS~12\,$\mu$m flux has also been scaled by the same
factor. 
No correction
has been applied for the WISE 23-$\mu$m photometry assuming
that emission at long wavelengths is dominated by
the emission from \#1.

\begin{figure*}
\centering
\includegraphics[width=9.1cm]{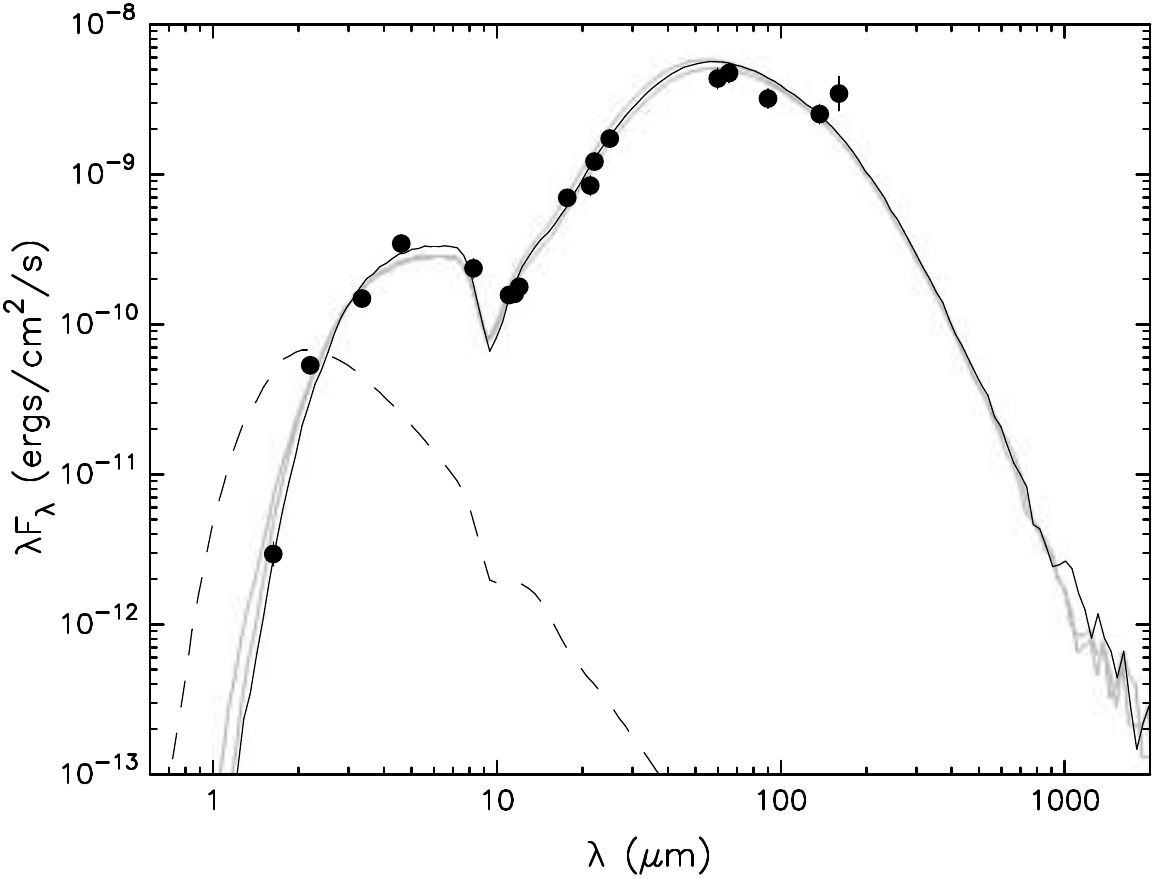}
\includegraphics[width=9.1cm]{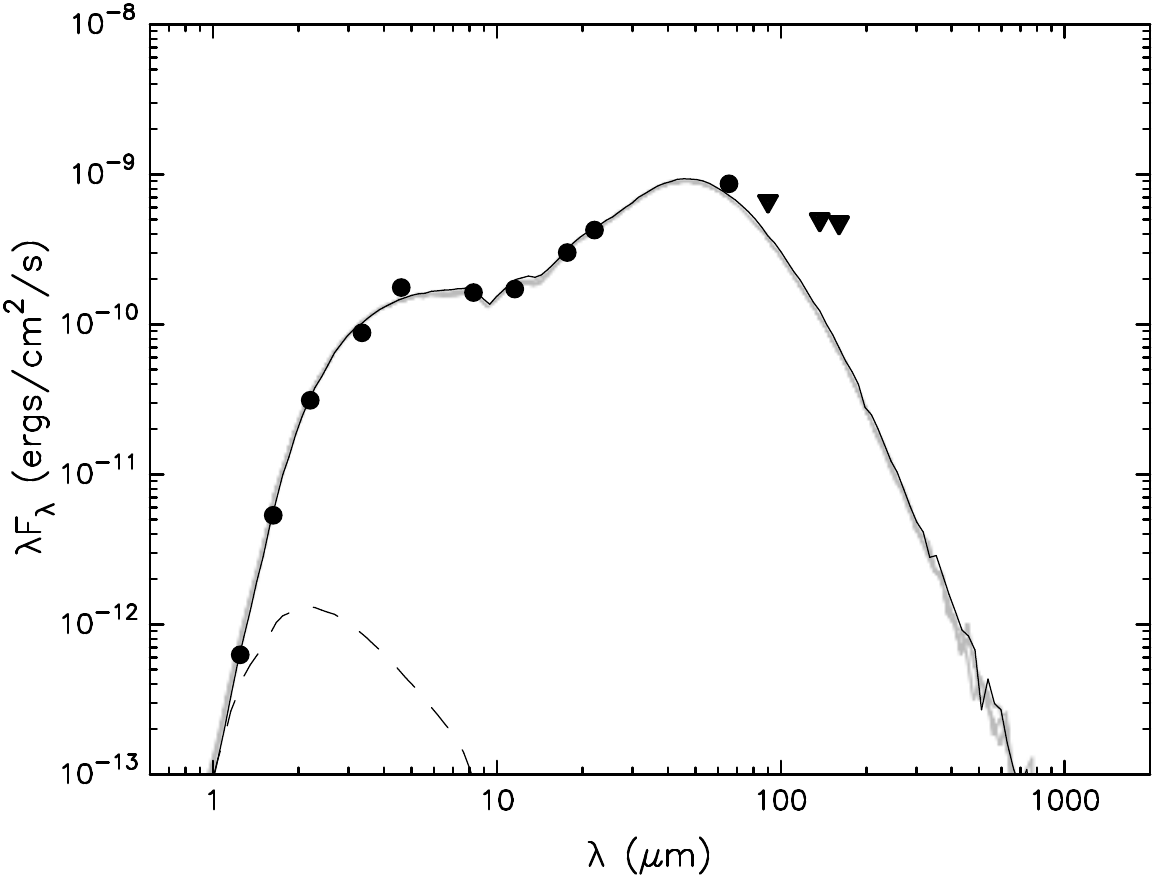}
\caption{The left panel shows the SED of \#1.
The filled circles show the data from WFCAM at $H$ and $K$,
MSX at 8.28 and 21.34\,$\mu$m, Michelle at 11.2\,$\mu$m, WISE at 3.6--22\,$\mu$m,
IRAS at 12--60\,$\mu$m, and AKARI at 18--140\,$\mu$m wavelengths.
The right panel shows the SED of \#8.  The filled circles show the data from WFCAM at $J$, $H$ and $K$,
MSX at 8.28\,$\mu$m, Michelle at 11.2\,$\mu$m, WISE 3.6--22\,$\mu$m and
AKARI at 18 and  65\,$\mu$m; the
downward directed triangles show the AKARI 90--160\,$\mu$m-band data, which are
treated as upper limits only. The continuous line
shows the best fit model and the grey lines show subsequent good
fits for ($\chi^2-\chi^2_{best fit}$) per data point $<$ 3.
The dashed line corresponds to the stellar photosphere for the
central source of the best fitting model, as it would look
in the absence of circumstellar dust (but including interstellar
extinction).}
\label{07422_sed}%
\end{figure*}

The flux in MSX band A 
includes emission from the rest of the cluster.  MSX flux 
in band A is corrected by the relative contribution from 
source \#1 towards the observed flux, which is the emission 
from the whole cluster.  This correction factor is derived 
by interpolating for the value at 8.28\,$\mu$m between the
relative contribution of \#1 in the WISE 4.6-$\mu$m data
and in the Michelle 11.2-$\mu$m data.  No corrections are
applied to the MSX band E data.  MSX fluxes of \#1 in bands 
C and D are not used in our fit since these are of poor quality.

The IRAS~100\,$\mu$m flux of this region is of moderate quality
only and will be contaminated by additional emission from the
cluster at the low spatial resolution of IRAS.  It is also
affected by IR cirrus.  So it is not included 
in the model fit.  

The left panel in Fig. \ref{07422_sed} shows the model 
fit to the SED of source \#1.  Column 2 of
Table \ref{tab:results} shows the results of the SED
modelling for source \#1.  In addition to the best
fit model, there were only two additional models with 
($\chi^2-\chi^2_{best fit}$) per data point $<$ 3.
Out of those, one differs from the best fit model only in 
foreground and circumstellar A$_V$ and the angle of inclination of the
disk axis. Hence, the errors given against the parameter 
values in Table \ref{tab:results}  
are the differences in the parameter values of the best
fit model and of the additional fit.

We derived a mass, luminosity and temperature of 
10.7\,M$_{\odot}$, 3.9$\times$10$^3$\,L$_{\odot}$ 
and 7840\,K respectively from the fit.  \#1 is actively 
accreting mass and it is likely to be a Class-I protostar.
A foreground visual extinction A$_V$=16\,mag is obtained
from the model fit, whereas Fig. \ref{07422JHKcol} implies 
a much larger A$_V$ for the IR source.  This suggests that 
\#1 has a large amount of circumstellar matter, probably 
as a disk.  This is implied by an  A$_V$(circumstellar)=25.4\,mag
given by the fit. 
High angular resolution observations in the 
far-IR to sub-mm wavelengths are required to further
constrain the model fit.

\subsubsection{Source \#8}
\label{sed_8}

WFCAM $JHK$, AKARI-IRC and FIS, WISE and MSX band-A data 
are used for fitting the SED of \#8.  MSX C, D and E-band 
data are not included in the final fit since they are of 
poor quality; (nevertheless, they agree well with the 
current fit and including them in the fit did not change 
the fit results).  WFCAM $JHK$ photometry are obtained in
a 1.5\arcsec-diameter aperture to reduce the contamination
from the neighbours and the nebulosity in the cluster to 
a minimum.  

The right panel in Fig. \ref{07422_sed} 
shows the model fit to the SED of \#8.  Table \ref{tab:results} 
shows the parameters of the best fit model.  Upon the 
initial attempts to fit the data, it was noticed that 
AKARI-FIS data beyond 65\,$\mu$m exhibit excess with respect
to the model fit, which systematically increases towards longer
wavelengths.  Hence the AKARI 90, 140 and 
160-$\mu$m data were not included in the fit and 
were treated as upper limits with low probability only.  
These data points are are shown as downward directed 
triangles in the figure.  The data points shortward of 
65\,$\mu$m fit the model well; a stellar mass, temperature, 
luminosity and age of 6.7\,M$_{\odot}$, 19950\,K, 
1.4$\times$10$^3$\,L$_{\odot}$ and 6.0$\times$10$^5$\,yrs 
respectively are derived.  \#8 appears to be older
and much less embedded than \#1. The model fit yields
a foreground extinction similar to that for \#1, but
a much lower circumstellar extinction.
The high temperature of \#8 suggests that it might have 
formed an H{\sc{ii}} region around it.  However, note 
that even after 
formation of an H{\sc{ii}} region, the young star can still
grow through accretion of ionized gas (eg. Keto \cite{keto07}).

Nevertheless, it should be noted that the temparature of
\#8 derived by the model fit (19950\,K) is too high even for 
a main sequence star of its derived mass of 6.7\,M$_{\odot}$.
This can happen if there is extra contribution to the flux 
observed at short wavelengths which may make the object appear 
hotter. It should also be noted that a circumstellar extinction 
of A$_V$=0.64\,mag is abnormally low for a young source 
exhibiting a large IR excess as seen in Fig. \ref{07422JHKcol}.
It is very much possible that either an angle of 
49.5$^{\circ}$ estimated by the SED modelling is wrong
or more than one source is contributing to the near-
and mid-IR fluxes that we use for model fit.  
High angular resolution observations are required
for \#8 to understand the real nature of the
dominant YSO here.

The deviations of the AKARI-FIS data from the model fit,
which increase towards longer wavelengths, suggest an extra
contribution in the beam at longer wavelengths from a 
source colder than \#8.  The feature labelled ``Blob''
(Figs. \ref{wise3} and \ref{wise123} and \S\ref{CII})
needs to be studied at better spatial resolution
and higher sensitivity at longer wavelengths to see if
the it hosts a source much younger than \#8 and if
that is the one contributing to the emission in the
FIS bands, which causes the FIS fluxes to deviate
systematically from the model fit.
Some other near-IR
sources  like \#9, also need to be understood 
in detail.

\begin{table*}
\caption{Results from SED fitting}             
\label{tab:results}      	
\centering                      
\begin{tabular}{lll}        	
\hline\hline					\\[-2mm]
Parameter                       &\multicolumn{2}{c}{Best fit values$^{\mathrm{a}}$}				\\
				&Source \#1				&Source \#8				\\
\hline              												\\[-2mm]         
Stellar mass (M$_{\odot}$)      &10.7 (+0, -0.2)			&6.7  					\\
Stellar age (yr)                &3.0 (+0, -0.2)$\times$10$^{4}$		&6.0$\times$10$^{5}$			\\
Stellar radius (R$_{\odot})$	&33.6 (+4.5, -0)			&3.1					\\
Stellar temperature (K)		&7840 (+0, -625)			&19950					\\
Disk mass (M$_{\odot}$)         &1.3 (+0, -0.3)$\times$10$^{-2}$	&3.6$\times$10$^{-3}$			\\
Disk accretion rate (M$_{\odot}$yr$^{-1}$)&2.9 (+0, -2.42)$\times$10$^{-6}$ &7.6$\times$10$^{-7}$		\\
Disk/envelope inner radius (AU) &8.4 (+2.2, -0)				&12.7					\\
Disk outer radius (AU)		&16.1 (+1.2, -0)			&134					\\
Envelope mass (M$_{\odot}$)     &6.1 (+0.5, -0)$\times$10$^{2}$ 	&2.3					\\
Envelope accretion rate (M$_{\odot}$yr$^{-1}$)  &9.3 (+1.1, -0)$\times$10$^{-4}$ 	&7.5$\times$10$^{-7}$	\\
Envelope outer radius (AU)      &1.0$\times$10$^{5}$ 			&4.3$\times$10$^{4}$			\\
Total Luminosity (L$_{\odot}$)  &3.9 (+0.3, -0)$\times$10$^{3}$	&1.4$\times$10$^{3}$				\\
Angle of inclination of the disk axis$^{\mathrm{b}}$ ($^{\circ}$) &31.8 ($\pm$18.5)	&49.5 (+14.6, -8.1)	\\
A$_V$ (Foreground)		&16.1 (+4.7, -0.8)			&17.1 (+0, -0.5)			\\
A$_V$ (Circumstellar)           &25.4 (+0, -8.7)   			&0.6 (+0, -0.1)				\\

\hline 
\end{tabular}
\begin{list}{}{}
\item[$^{\mathrm{a}}$] The values given in parenthesis are the differences
in the parameters of the additional models with
($\chi^2-\chi^2_{best fit}$) per data point $<$ 3, from those 
of the best fit model. No errors are given when the parameters 
of these models do not differ; nevertheless, the uncertainties could
be higher as described in \S\ref{fitting}.
A distance of 4.01\,kpc is used.
\item[$^{\mathrm{b}}$] The angle of inclination of the disk axis is with respect to the line of sight.
\end{list}
\end{table*}

\subsection{The general picture massive star formation in IRAS~07422 
as seen in the infrared}

At a distance of 4\,kpc, the cloud detected in the WISE images
is larger than $\sim$17.5\,pc$\times$35\,pc.  Most of the star 
forming activity seems to be happening in a region of diameter
$<$11.5\,pc in the southern
region of the cloud.  The four clusters/cluster candidates ``CI--CIV'' detected by 
us are in this region.  We see H$_2$ line emission features which 
are likely to be caused by jets from ``CI--CIII''.  The brightest
object in ``CIV'' also has infrared colours typical of YSOs.
At least in clusters ``CI--CIII'', the brightest object in near-
and mid-IR seems to be located near the centres of the clusters.

Several theoretical and observational studies of 
massive star forming regions suggest that massive 
star formation is intimately linked to the hierarchical 
formation of massive stellar clusters, where massive stars 
form near the centres of massive sub-clusters and grow by 
competitive accretion aided by the cluster potential.  
In this scenario, several subclusters form in a massive 
clump, massive stars near the central regions of these 
subclusters grow through competitive accretion and finally
the subclusters merge (eg. Testi et al. \cite{testi00},
Clarke, Bonnell \& Hillenbrand \cite{clarke00}, Bonnell, Bate
\& Vine \cite{bonnell03}, Bonnell \& Bate \cite{bonnell06}, 
Bonnell \& Smith \cite{bonnell11}).  Our observations reported
here support this scenario.   (However, note that 
precise radial velocity studies to understand the
dymamics of the gas associated with this region 
are required for a confirmation).
At least in ``CI'' and
``CII'', the brightest objects in the infrared are seen
close to the centres of the clusters hosting them.  In
``CIII'' also, the brightest object seems to be located
close to the centre of the elongated pattern of distribution
of the reddened objects. WISE observations, though deep, 
lack the spatial resolution to resolve the members of the 
clusters at these large distances.  More sensitive 
$K$-band and thermal IR imaging with high angular resolution 
are required to understand the census and distribution of 
the young stars in these clusters.  High angular 
resolution observations at sub-mm and mm wavelengths 
are also required to understand if there are younger 
members in these clusters than those we see at near-IR
and to get a more detailed picture of the star and cluster
forming activity in the cloud.

\section{Conclusions}

\begin{enumerate}
\item At least two (and possibly four) young subclusters,
embedded in a cloud are discovered in IRAS~07422.
Narrow-band H$_2$ 1-0 S(1) line images reveal eight or 
more outflows from them.

\item The most prominent and youngest cluster in this region 
is ``CI'', associated with the IRAS source and the
brightest of the MSX and AKARI sources. Multiple star formation
is happening in ``CI''; H$_2$ line emission
features observed in its vicinity are suggestive
of at least six outflows produced by YSOs in ``CI''.
\#1 is the most luminous object in this region. We derive a mass
and luminosity of 10.7\,M$_{\odot}$ and
3.9$\times$10$^3$\,L${\odot}$ respectively by modelling its SED. 
This object appears 
to be very young and and is probably a Class-I protostar. High
angular resolution observations at 11.2\,$\mu$m reveal three
of the embedded sources in ``CI''.

\item Source \#1 in ``CI'' is located close to an
extinction lane, inclined at $\sim$53$^{\circ}$ east of north.
A set of three H$_2$ knots are seen aligned in a direction
roughly perpendicular to the extinction lane.  High angular
resolution at mm wavelengths are reqired to understand if
the extinction lane is a circumstellar disk seen nearly
edge on.

\item \#8 is the brightest object in the near-IR, located in the 
second most prominent cluster in this region, ``CII''.  
The second brightest object 
in the AKARI, WISE  and MSX appears to be associated with 
this region, but the spatial resolution of any of these
surveys is not sufficient to identify between \#8 and any of
its very close neighbours.  There is an outflow detected from
this cluster.  It is not clear if \#8 or any of it very
close neighbours is responsible for the outflow.
Modelling the SED of \#8 yields a mass, luminosity and
temperature of 6.7\,M$_{\odot}$, 1.4$\times$10$^3$\,L${\odot}$
and 19950\,K respectively.  This temperature is high 
even for a main sequence star of 6.7\,M$_\odot$ suggesting
possible extra contributions to the flux at short wavelengths.
The SED fit shows the four AKARI-FIS fluxes to be deviating
systematically from the fit, with the residuals increasing
towards longer wavelengths.  These are suggestive of contribution
from a colder component to the flux measured by AKARI.
Cluster candidates ``CIII'' and ``CIV'' also host YSOs;
``CIII'' appears to host one or more outflows.

\item  Multiple star formation is taking place inside
more than one sub-cluster within a cloud in this region.
This region appears to be a case of hierarchical mode of
cluster formation where massive star formation
takes place in multiple sub-clusters in a cloud and the massive
stars at the centres of the sub-clusters grow through competitive
accretion.

\item Observations at IR through mm wavelengths at better 
angular resolution are required for a more detailed understanding
of the star formation activity in this cloud.

\end{enumerate}

\begin{acknowledgements}
The UKIRT is operated by the Joint Astronomy Centre on behalf 
of the Science and Technology Facilities Council (STFC) of the 
UK.  The UKIRT-WFCAM data presented in this paper are obtained 
during the UKIDSS back up time. I thank the Cambridge Astronomical 
Survey Unit (CASU) for processing the WFCAM data, and the WFCAM 
Science Archive (WSA) for making the data available.  The 
UKIRT-Michelle data are downloaded from the UKIRT archive.  
This work makes use of data obtained with AKARI, a JAXA project with
the participation of ESA, downloaded from CDS, Strasbourg, France.
We also use data products from the Wide-field 
Infrared Survey Explorer, which is a joint project of the 
University of California, Los Angeles, and the Jet Propulsion 
Laboratory/California Institute of Technology, funded by the 
National Aeronautics and Space Administration.  WISE, IRAS and
MSX images and photometry were downloaded from the NASA/IPAC 
Infrared Science Archive.  I thank Jack Ehle and Thor Wold for
carrying out the WFCAM observations.  I also thank the 
referee Thomas Stanke and the editor Malcolm Walmsley for 
their comments and suggestions which have improved the
quality of the paper.
\end{acknowledgements}

\end{document}